\documentclass[aps, prb, twocolumn, showpacs, 10pt, floatfix, superscriptaddress, floatfix, longbibliography]{revtex4-2}

\usepackage{setspace}
\usepackage{amsthm}

\usepackage[english]{babel}
\usepackage[utf8]{inputenc}
\usepackage[T1]{fontenc}

\usepackage{amsmath}
\usepackage{braket}
\usepackage{amssymb}
\usepackage[normalem]{ulem}
\usepackage{dsfont}
\usepackage{bm}
\usepackage[dvipsnames]{xcolor}
\usepackage{stmaryrd}
\usepackage{bbm}
\usepackage{mathtools}

\usepackage{multirow}
\usepackage{enumitem}

\usepackage{color}
\usepackage[dvipsnames]{xcolor}

\usepackage[most]{tcolorbox}
\newtcbox{\othermathbox}[1][]{nobeforeafter, math upper, tcbox raise base, enhanced, rounded corners, colback=black!5, colframe=black, left=0.3em, top=0.3em, right=0.3em, bottom=0.4em}
\usepackage{empheq}

\usepackage{hyperref}
\hypersetup{
    colorlinks=true,
    citecolor=Green,
    linkcolor=Maroon,
    urlcolor=Blue
    }

\makeatletter
\def\l@subsection#1#2{}
\def\l@subsubsection#1#2{}
\makeatother

\usepackage{lipsum}  

\newcommand{\beq}{\begin{equation}}
\newcommand{\eeq}{\end{equation}}

\definecolor{MyRed}{RGB}{220,60,10}
\definecolor{LightYellow}{RGB}{255,245,108}
\definecolor{LightBrown}{RGB}{255,188,0}
\definecolor{MiddleBrown}{RGB}{199,146,0}
\definecolor{DarkBrown}{RGB}{143,104,0}
\definecolor{DarkerBrown}{RGB}{87,62,0}
\definecolor{Purple}{RGB}{255,0,188}

\begin{document}

\title{Entanglement transitions in structured and random nonunitary Gaussian circuits}

\author{Bastien Lapierre}
\email{blapierre@princeton.edu}
\affiliation{Department of Physics, Princeton University, Princeton, New Jersey, 08544, USA}

\author{Liang-Hong Mo}
\email{lm8598@princeton.edu}
\affiliation{Department of Physics, Princeton University, Princeton, New Jersey, 08544, USA}

\author{Shinsei Ryu}
\email{shinseir@princeton.edu}
\affiliation{Department of Physics, Princeton University, Princeton, New Jersey, 08544, USA}

\date{\today}

\begin{abstract}
We study measurement-induced phase transitions in quantum circuits consisting of kicked Ising models with postselected weak measurements, whose dynamics can be mapped onto a classical dynamical system. For a periodic (Floquet) non-unitary evolution, such circuits are exactly tractable and admit 
volume-to-area law transitions.
We show that breaking time-translation symmetry down to a quasiperiodic (Fibonacci) time evolution leads to the emergence of a critical phase with tunable effective central charge and with a fractal origin. 
Furthermore, for some classes of random non-unitary circuits, we demonstrate the robustness of the volume-to-area law phase transition for arbitrary random realizations, thanks to the emergent compactness of the classical map encoding the circuit's dynamics.
\end{abstract}


\maketitle

\section{Introduction}

The nonequilibrium dynamics of closed quantum many-body systems can give rise to phenomena and phases that have no equilibrium counterparts~\cite{Else2016, PhysRevLett.116.250401, PhysRevX.7.011026, PhysRevX.3.031005, Wang_2013, PhysRevLett.99.220403, Zhang:2017bwh, Kyprianidis_2021}. 
Nevertheless, such isolated dynamics is often challenging to realize due to the presence of dissipation and decoherence in experiments.
It was recently appreciated that the interplay between unitary and non-unitary time evolution, the latter stemming from dissipation or measurements, may be used to stabilize new types of phase transitions. 
One prominent example is measurement-induced phase transitions (MIPTs)~\cite{PhysRevX.9.031009,PhysRevB.98.205136, PhysRevB.100.134306, PhysRevB.99.224307, PhysRevB.100.064204, PhysRevX.10.041020, bao2020theory,PhysRevB.101.104302,Noel_2022} that arise from the interplay between measurements and unitary evolution. MIPTs are witnessed by quantum information-theoretical quantities, such as entanglement entropy, and separate a quantum Zeno-like area law phase and a volume law phase, through an alternation between local measurements and unitary evolution. 
While original setups focused on Haar random unitaries as well as Clifford circuits,
it was understood that Gaussian circuits, which are built out of free-fermionic operators, may also display similar entanglement phase transitions~\cite{10.21468/SciPostPhys.7.2.024, PhysRevResearch.2.033017, PhysRevLett.126.170602, PhysRevB.106.134206, Turkeshi_2021,Le_Gal_2023,PhysRevB.106.L220304,PhysRevLett.130.230401,PhysRevResearch.6.013131}.

It was shown in~\cite{PhysRevLett.130.230401} that transitions from area-to-volume law phases of the entanglement entropy may be studied exactly in Gaussian non-unitary circuits with spatial translation invariance, through the requirement that the circuit is periodic in time.
Such a requirement enables an analytical derivation of the critical exponent across the transition, which is described by a critical theory with zero effective central charge~\cite{PhysRevLett.126.170602, PhysRevB.103.224210}.
It is, however, crucial to understand whether such analytically tractable MIPTs are fine-tuned because of time-translation symmetry, and it is thus natural to study their robustness in the absence of such symmetry.

\begin{figure}[t]
	\includegraphics[width=0.48\textwidth]{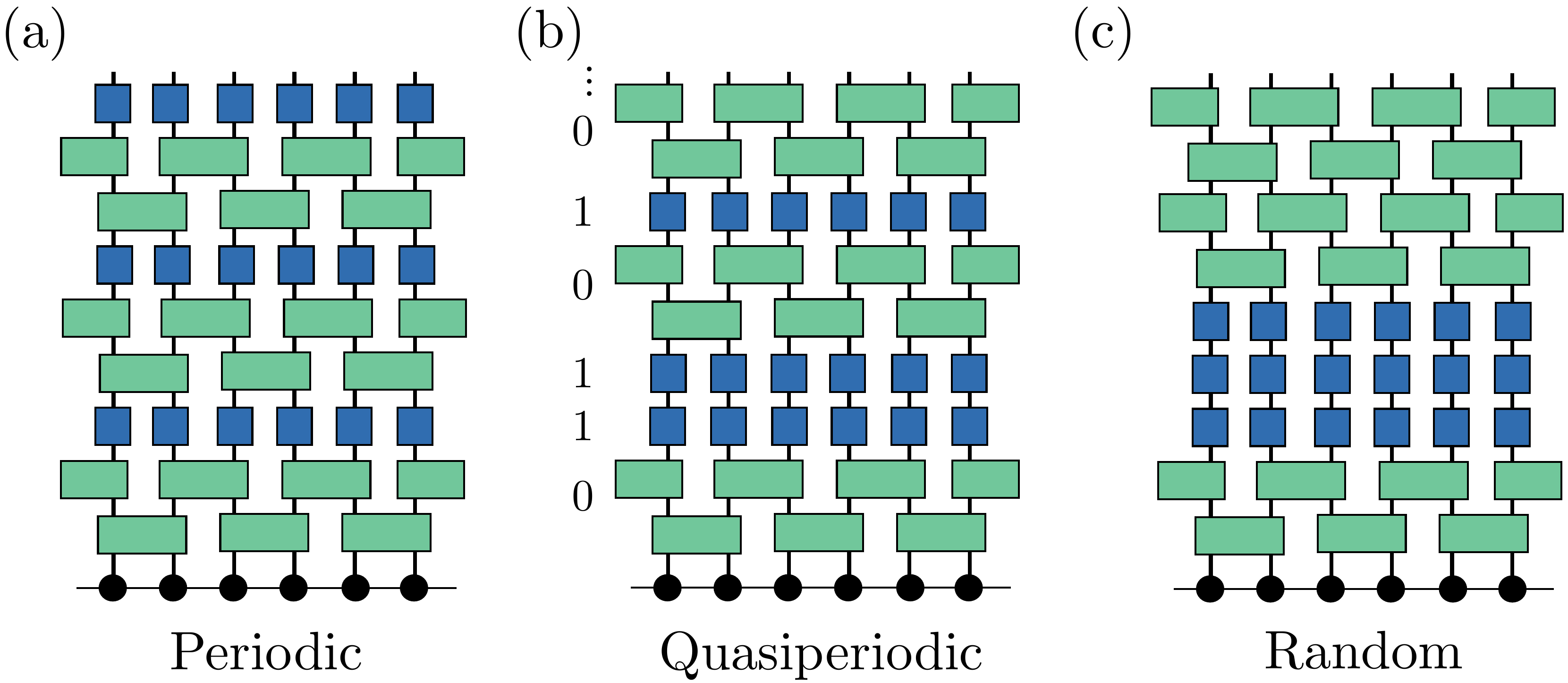}
	\caption{\label{fig:periodquasirando} Illustration of the three different types of non-unitary circuits studied in this work, constructed from the gates~\eqref{unitarygates}, each corresponding to a different degree of time-translation breaking. (a) A periodic (Floquet) circuit, discussed in Sec.~\ref{sec:periodiccircuit}.  (b) A quasiperiodic circuit whose non-unitary evolution is determined by a Fibonacci word `011010...', discussed in Sec.~\ref{sec:quasiperiodiccircuit}. (c) A random circuit built by alternating randomly between two non-unitary gates, discussed in Sec.~\ref{sec:randomcircuit}. Our goal will be to understand the different entanglement phases in these three different scenarios.}
\end{figure}

In this work, we analytically and numerically study the fate of the volume-to-area law transition in a class of Gaussian non-unitary circuits that are either quasiperiodic or random in time, and we relate the different entanglement phases of such circuits to localization transitions in disordered (quasiperiodic or random) tight-binding models.  
We first focus our attention on the Fibonacci quasiperiodic sequence, which has appeared prominently in the context of quasicrystals~\cite{RevModPhys.93.045001, Damanik_2016}, and more recently in driven quantum many-body systems~\cite{PhysRevLett.120.070602, Dumitrescu_2022,PhysRevResearch.2.033461, PhysRevResearch.3.023044, schmid2024selfsimilarphasediagramfibonaccidriven, PhysRevLett.131.250401}. 
In both of these settings, the quasiperiodicity leads to rich physics, including, e.g., multifractal critical states for quasicrystals as well as fractal heating dynamics for quasiperiodically driven quantum systems~\cite{PhysRevResearch.2.033461, PhysRevResearch.3.023044}.
As we demonstrate, fractality similarly emerges in the entanglement dynamics of the Gaussian circuit and leads to an extended critical phase defined by a logarithmic growth of entanglement entropy with a tunable effective central charge.
Furthermore, we study the long-time dynamics of purely random non-unitary circuits, for which the time evolution of fermionic coherent states reduces to a random walk in SL$(2,\mathbb{C})$. Although on general grounds the volume law phase is expected to vanish due to Furstenberg's theorem, we analytically prove that a properly designed random circuit can exhibit exact and robust entanglement phase transitions between area and volume law phases. The origin of the transition is that for these circuits, there is an emergent SU$(2)$ dynamics for the coherent states that protects the volume law phase, even for a single random realization.

The rest of the paper is organized as follows. First, we introduce the circuit and set the stage for our calculations in Sec.~\ref{sec:setup}. Then, in Sec.~\ref{sec:periodiccircuit}, we consider the case of a periodic non-unitary circuit, following the approach discussed in~\cite{PhysRevLett.130.230401}. We then break time-translation invariance and study Fibonacci quasiperiodic circuits in Sec.~\ref{sec:quasiperiodiccircuit}, and highlight their fractal structures, in contrast to the periodic case. We finally discuss the case of a random non-unitary circuit in Sec.~\ref{sec:randomcircuit}, and discuss the existence of the entanglement phase transition for both fully random and `random dipolar' circuits.

\section{Setup of the non-unitary circuits}
\label{sec:setup}

In this paper, we consider a family of Gaussian circuits consisting of the following unitary gates, similar to a one-dimensional kicked Ising model on a chain of length $L$ with periodic boundaries,
\begin{align}
\begin{split}
\label{unitarygates}
U_X(t) &= e^{-it \sum_{j=1}^{L}X_{j}},\\ 
U_{YY}(t) &= e^{-it\sum_{j=1}^{L}Y_{j+1}Y_{j}},\\ 
U_{ZZ}(t) &= e^{-it \sum_{j=1}^{L}Z_{j+1}Z_{j}}.
\end{split}
\end{align}
It can be readily shown that all gates can be written in terms of free fermions using the Jordan-Wigner transformation. 
It was noted in~\cite{PhysRevA.104.062614} that fermionic coherent states
\begin{equation}
\label{coherentstate}
|\psi(A,f)\rangle = A\prod_{k\in\Lambda_+}(1+f(k)c_{k}^{\dagger}c_{-k}^{\dagger})|0\rangle,
\end{equation}
(where $\Lambda =\frac{2\pi}{L} \{-\frac{L}{2}+\frac{1}{2},\ldots,\frac{L}{2}-\frac{1}{2}\}$ and $\Lambda_+$ is restricted to the positive momenta)
have a remarkably simple time evolution under the gates~\eqref{unitarygates}. In fact, time evolution maps any such state to a new coherent state via a M\"obius transformation,
\begin{equation}
U_{\mathcal{O}}(t)|\psi(A,f(k))\rangle = |\psi(\widetilde{A},\tilde{f}(k))\rangle, \quad \tilde{f}(k) = \frac{af(k)+b}{cf(k)+d},
\end{equation}
with $\mathcal{O}=X, YY, ZZ$. The coefficients of the M\"obius transformation depend on the momentum as well as the time $t$. Their explicit form is~\cite{PhysRevA.104.062614}
\begin{align}
\tilde{f}(k) &= e^{4it}f(k),\\
\tilde{f}(k) &= \frac{[1+\tan^2(k/2)e^{4it}]f(k)+i\tan(k/2)(1-e^{4it})}{-i\tan(k/2)(1-e^{4it})f(k)+\tan^2(k/2)+e^{4it}},\nonumber\\
\tilde{f}(k) &= \frac{[1+\tan^2(k/2)e^{4it}]f(k)-i\tan(k/2)(1-e^{4it})}{i\tan(k/2)(1-e^{4it})f(k)+\tan^2(k/2)+e^{4it}},
\nonumber
\end{align}
for ${\cal O}=X, YY, ZZ$, 
respectively.
These M\"obius transformations can be recast into SU$(2)$ matrices after being properly normalized, such that applying multiple time steps amounts to compositions of SU$(2)$ transformations at the level of coherent states.
The non-unitarity of the circuit is implemented by $U_X(i\lambda)$, which is equivalent to coupling each site of the chain to an ancilla qubit and postselecting the outcome by fixing the same sign of $\lambda$ at each site and at each measurement step~\cite{PhysRevLett.130.230401}.
The associated M\"obius transformation is thus a scaling of the form $e^{-4\lambda}f(k)$.
As a consequence, composing the unitary time evolution via the gates ~\eqref{unitarygates} with the non-unitary operator $U_X(i\lambda)$ leads to an overall SL$(2,\mathbb{C})$ transformation of the initial coherent state.
In the case of a periodic circuit, as will be studied in Sec.~\ref{sec:periodiccircuit}, stroboscopic time evolution simply amounts to composing a one-cycle transformation $n$ times with itself.
Beyond periodic circuits, such as quasiperiodic circuits and random circuits studied in Sec.~\ref{sec:quasiperiodiccircuit} and~\ref{sec:randomcircuit} respectively, compositions of SL$(2,\mathbb{C})$ transformations are no longer periodic, which will in general spoil the solvability of the circuit.
Nevertheless, fruitful analogies with transfer matrices arising in one-dimensional disordered tight-binding models can be used to deduce the entanglement properties of the steady state, even in the absence of time-translation invariance.

In order to characterize the steady state entanglement of the circuit, we will study the scaling of von Neumann entanglement entropy, defined as
\begin{equation}
    S_A(\ell)=-\text{Tr}\, \rho_A\log\rho_A,
\end{equation}
for a subsystem $A$ of size $\ell$ in a given steady state $\rho=|\psi_{\infty}\rangle\langle \psi_{\infty}|$. By construction, the steady state assumes the form of a fermionic coherent state, characterized by $f_{\infty}(k)$, in the limit $\ell\ll n\ll L$, where $n$ is the number of applied gates.
Because the circuit is Gaussian, the entanglement entropy can be
computed from correlation matrices~\cite{Calabrese_2005, Peschel_2003}. 
More precisely, we introduce the correlation matrix in terms of Majorana operators $a_{2j}=i(c_j-c_j^{\dagger})$, $a_{2j-1}=c_j+c_j^{\dagger}$:
\begin{equation}
\langle \psi_{\infty}|(a_{2l-1}, a_{2l})^T(a_{m-1},a_{2m})|\psi_{\infty} \rangle = \delta_{lm}+i\Gamma_{lm}.
\end{equation}
In particular, the correlation matrix $\Gamma$ can then be written as a block Toeplitz matrix with coefficients given by the Fourier coefficients of $f_{\infty}(k)$, defined in the thermodynamic limit as~\cite{PhysRevLett.130.230401},
\begin{align}
\label{fouriercoeffcieitnseq}
\varphi_j &= \frac{i}{2\pi}\int_{-\pi}^{\pi}\text{d}ke^{-ikj}\frac{f_{\infty}(k)+f_{\infty}^*(k)}{1+|f_{\infty}(k)|^2},\\
\psi_j &= \frac{1}{2\pi}\int_{-\pi}^{\pi}\text{d}ke^{-ikj}\frac{f_{\infty}(k)-f_{\infty}^*(k)+|f_{\infty}(k)|^2-1}{1+|f_{\infty}(k)|^2}. \nonumber
\end{align}
The entanglement entropy is then related to the imaginary part of the eigenvalues $\{\pm\lambda_j\}_{j=1}^{\ell}$ of $\Gamma$ through
\begin{equation}
S_A(\ell) = \sum_{j=1}^{\ell}h_1(\text{Im}(\lambda_j)),
\end{equation}
with the function $h_1(x) = -\frac{1+x}{2}\log\frac{1+x}{2}-\frac{1-x}{2}\log\frac{1-x}{2}$.
In the following sections, we will consider different types of non-unitary circuits and deduce entanglement scaling from the steady-state properties.

\section{Periodic circuits}
\label{sec:periodiccircuit}

We first discuss the case of periodic non-unitary circuits, closely following the strategy presented in~\cite{PhysRevLett.130.230401}, where the circuit comprises unitary time evolution in terms of the gates $U_{ZZ}(t)$, $U_{YY}(t)$, $U_X(t)$, as well as postselected weak measurements represented by the non-unitary operator $U_X(i\lambda)$.

By restricting the dynamics of the problem to initial coherent states~\eqref{coherentstate}, we essentially map the dynamics of the periodic circuit to a classical dynamical system. 
As a consequence, the occurrence of a transition between area and volume law of entanglement entropy can be inferred from the classification of the underlying one-cycle M\"obius transformation, $\tilde{f}_1$. In fact, the evolution after $n$-cycles is simply $\tilde{f}_1\circ \tilde{f}_1 \circ \cdots\circ \tilde{f}_1$. The fixed point structure of this classical dynamical system depends on the trace squared of the associated SL$(2,\mathbb{C})$ matrix $M_{\text{F}}$, corresponding to the one-cycle Floquet operator $U_{\text{F}}$. If $0<\text{Tr}(M_{\text{F}})^2<4$, the transformation is said to be elliptic, and no fixed point appears in the dynamics. However, if $\text{Tr}(M_{\text{F}})^2>4$, the transformation is hyperbolic and a pair of unstable and stable fixed points emerge. The transition between the two is the parabolic transformation $\text{Tr}(M_{\text{F}})^2=4$, where both fixed points merge. If the trace squared is non-positive, the transformation is said to be loxodromic. As we will now see, the volume law phase emerges when an interval of momenta is associated to elliptic M\"obius transformations, while the area law appears when the transformation is hyperbolic for all momenta.
We note that similar dynamics appear in periodically driven critical systems, where the classification of one-cycle Möbius transformations provides different dynamical phases, leading to different growth regimes of the entanglement entropy $S_A(t)$~\cite{wen2018floquet, PhysRevResearch.2.023085, PhysRevB.102.205125}.

For concreteness, we choose the building blocks of the periodic circuit to be
\begin{align}
\label{eq:setpquasi}
U_0 &= U_{ZZ}(-T)U_X(i\lambda)U_{YY}(T),\\
U_1 &= U_{ZZ}(T)U_X(i\lambda)U_{YY}(-T),\nonumber
\end{align}
where the minus signs are interpreted as switching the coupling of the Hamiltonians. The Floquet non-unitary operator is defined as \footnote{We note that simpler choices of one-cycle operator, such as $U_{\text{F}}=U_0$ or $U_{\text{F}}=U_1$, lead to area to volume law transitions as well, but we nonetheless find it convenient to use~\eqref{eq:floquetunitary} in order to compare the phase diagrams in the next sections.}
\begin{equation}
\label{eq:floquetunitary}
U_{\text{F}} = U_0U_1.
\end{equation}
In this case, it can readily be shown that the associated one-cycle M\"obius transformation is real for any momenta.
In fact, the trace reads
\beq
\begin{aligned}
\text{Tr}(M_{\text{F}}) &= \frac{1}{8}[-3 + 13 \cosh(4 \lambda) + 
   2 \cosh(2 \lambda)^2 \\&\quad \times(4 \cos(4 T) - \cos(8 T) + 
      8 \cos(4 k) \sin(2 T)^4)].
\end{aligned}
\eeq
As found in~\cite{PhysRevLett.130.230401}, a transition between area and volume law scaling of entanglement entropy corresponds to a transition between elliptic and hyperbolic orbits of the associated M\"obius transformation. As long as there exists an extended interval of momenta such that the transformation is elliptic ($|\text{Tr}(M_{\text{F}})|<2$), a volume law emerges in the steady state, while the system follows area law if the transformation is hyperbolic ($|\text{Tr}(M_{\text{F}})|>2$) for all momenta. The phase boundary between these two cases can be found in the extremal case $k=\frac{\pi}{4}$, which leads to the following parametric equation in the $(T,\lambda)$ plane:
\begin{equation}
\label{eq:phaseboundwry}
 - 2 [-4 \cos(4 T) + \cos(8 T)] \cosh(2 \lambda)^2 + 
   5 \cosh(4 \lambda)=11.
\end{equation}
\begin{figure}[t]
\includegraphics[width=0.27\textwidth]{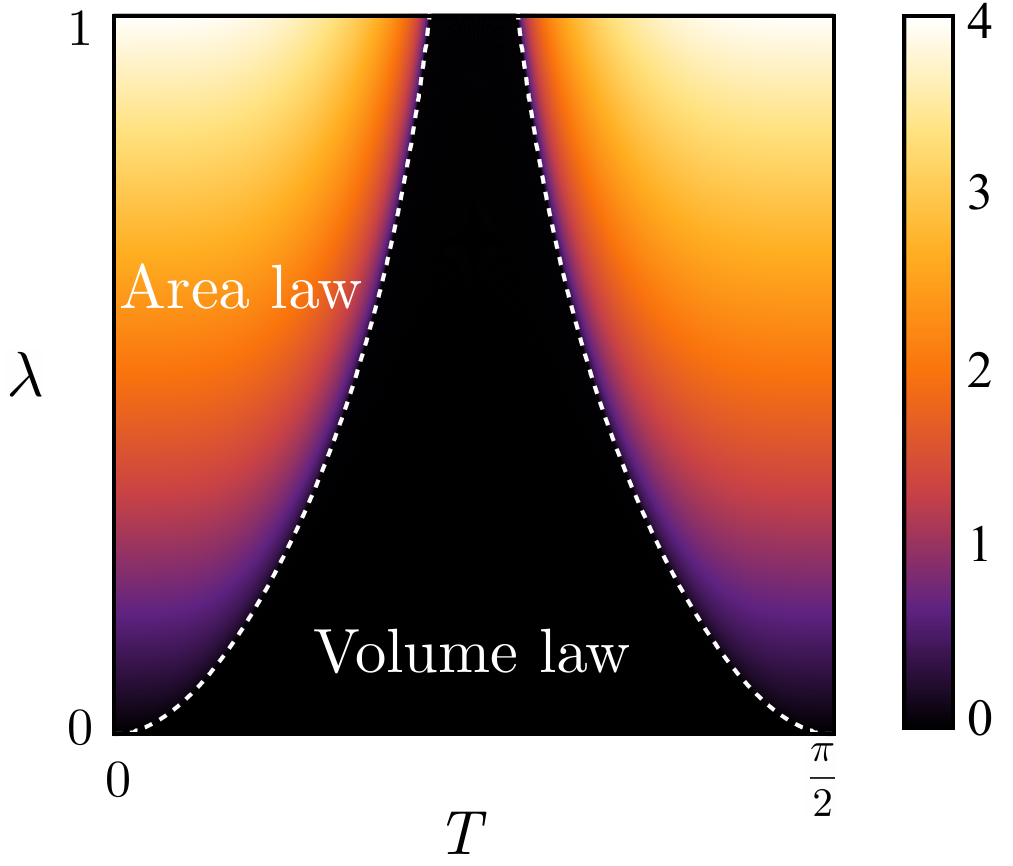}	\caption{\label{fig:phasediagramperiodic} Phase diagram for the periodic circuit defined in~\eqref{eq:floquetunitary}. The minimum Lyapunov exponent $\text{min}_{k\in\Lambda_+}(\lambda_L)$ is plotted, and takes a nonzero value in the area law as all momenta have nonzero exponent, while there exists an interval of momenta for which it is zero in the volume law phase. The phase boundary is determined by the parametric equation~\eqref{eq:phaseboundwry} (white dashed curve). The phase diagram is periodic in the $T$ direction with period $\frac{\pi}{2}$. }
\end{figure}
The associated phase diagram is illustrated in Fig.~\ref{fig:phasediagramperiodic}. As expected, increasing the measurement rate $\lambda$ leads to a progressive expansion of the area law region, which eventually fully dominates in the limit $\lambda\rightarrow\infty$, apart from the critical line at $T=\frac{\pi}{4}$. In particular, the phase diagram is $\frac{\pi}{2}$-periodic in $T$ since the unitary time evolution with the gates~\eqref{unitarygates} is itself periodic.
Equivalently, the different entanglement phases can be distinguished from the Lyapunov exponent, defined as
\begin{equation}
\lambda_L = \lim_{n\rightarrow\infty}\frac{1}{n}\log||M_{\text{F}}^n||,
\label{lyapunovexponent}
\end{equation}
where the norm is chosen to be the Frobenius norm.
The Lyapunov exponent may be expressed directly from the trace of the one-cycle SL$(2,\mathbb{C})$ transformation through
\begin{equation}
\lambda_L = \frac{1}{p}\log\left|\frac{|\text{Tr} (M_{\text{F}})|+\sqrt{|\text{Tr} (M_{\text{F}})|^2-4}}{2}\right|,
\label{lyapunovexponent2}
\end{equation}
where $p$ is the number of individual gates constituting one cycle. As a consequence, if the associated transformation is elliptic or parabolic, we conclude that the Lyapunov exponent vanishes, while it takes a finite nonzero value for any hyperbolic transformation. 
This will prove to be a convenient numerical tool when generalizing to quasiperiodic or random circuits, where the time evolution cannot be simply obtained by iterating a one-cycle transformation.

\begin{figure}[t]
	\includegraphics[width=0.48\textwidth]{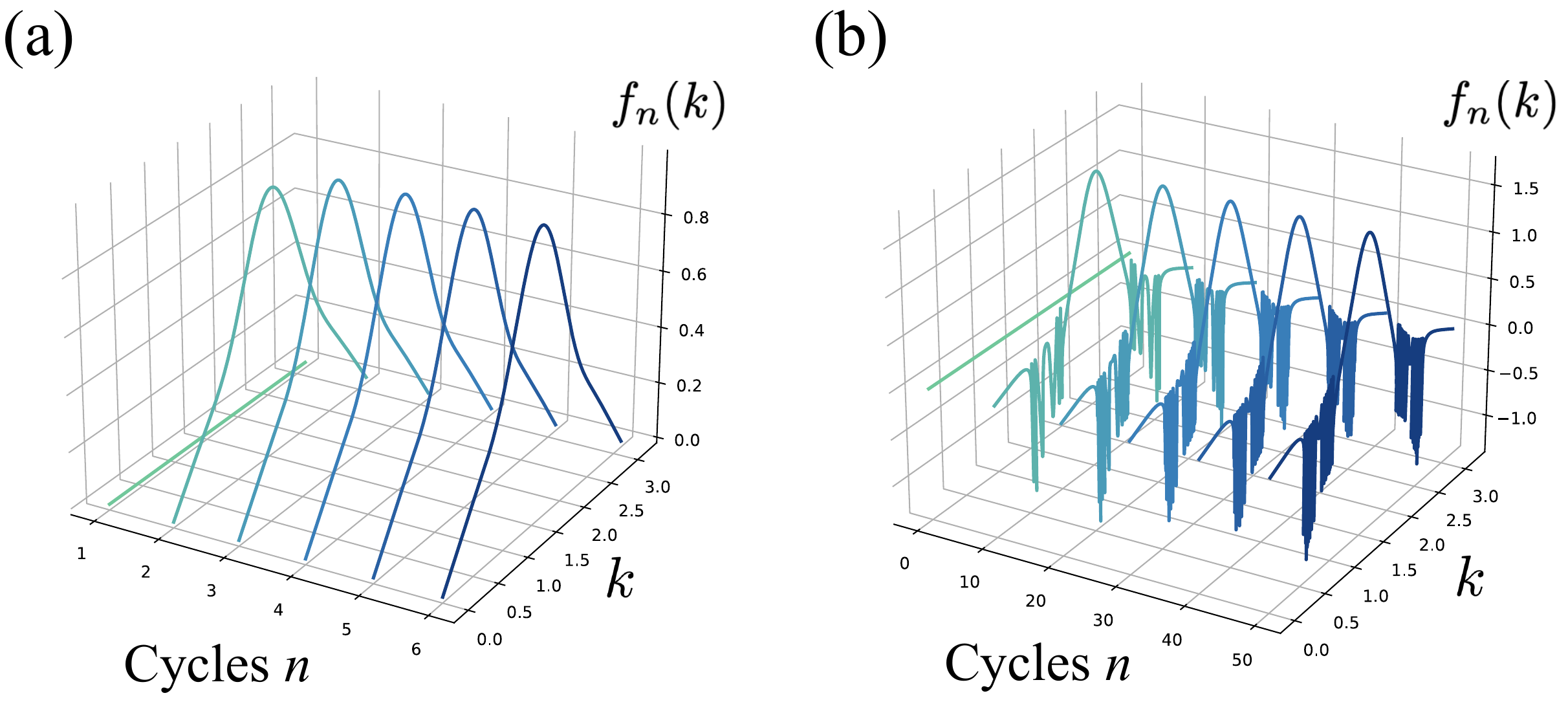}
	\caption{\label{fig:volumevsarea} Iterated momentum amplitude $f_n(k)$ for an initial state with all spins pointing in the $x$-direction, i.e., $f(k)=0$, in (a) the area law phase and in (b) the volume law phase. The emergence of intervals of momenta that are nonconvergent in the volume law phase is manifest. }
\end{figure}

In the area law phase, the M\"obius transformation converges to a fixed point for all $k$ such that the function $f_n(k)$ exponentially converges to a steady state characterized by a smooth function $f_{\infty}(k)$ in the infinite time limit, as seen on Fig.~\ref{fig:volumevsarea}(a) when starting from $f(k)=0$. As a consequence, the Fourier modes of $f_{\infty}(k)$ decay exponentially fast, which leads to the imaginary part of the eigenvalues of $\Gamma$ to be one, i.e., the entanglement entropy is independent of the subsystem size, $S_{A}\sim O(1)$.
On the other hand, in the volume law phase, there is an extended interval of momenta $k$ for which there is no convergence in the infinite time limit. The transformation $f_n(k)$ is non-smooth in this interval, Fig.~\ref{fig:volumevsarea}(b), leading to a slower decay of the Fourier coefficients. In fact, by averaging over fast oscillations, one can show that eigenvalues of the correlation matrix $\Gamma$ are smaller than one, and a non-zero linear coefficient arises in the entanglement entropy, $S_A\sim O(\ell)$.

The phase transition between area and volume law phases happens when there is a single momentum $k_c$ 
for which the one-cycle transformation is non-hyperbolic. In this case, there is a single discontinuity in $f_n(k)$ as $n\rightarrow \infty$, which implies power law decay of the Fourier coefficients, implying in turn a critical scaling $O(\log(\ell))$ of the entanglement entropy. In the model~\eqref{eq:floquetunitary}, this happens at $k_c=\frac{\pi}{4}$, for which one has that the one-cycle M\"obius transformation is parabolic. Because the transformation is parabolic, there is a critical point in the dynamics (i.e., a merger of the stable and unstable fixed points), which implies that $f_n(k_c)$ still converges in the infinite time limit. This implies that the $O(\log(\ell))$ term vanishes, which is to say that the effective central charge is zero. A nonvanishing $O(\log(\ell))$ term can appear if there is at least one critical momentum $k_c$ for which the one-cycle transformation is elliptic instead of parabolic.
Furthermore, as we will later study, a nonzero logarithmic coefficient may arise from having infinitely many \textit{isolated} critical momenta.

We stress that the existence of the volume-to-area law transition for this class of Gaussian circuits can be inferred from a purely algebraic perspective.
In fact, the entanglement transition is intrinsically linked to the non-compactness of the group SL$(2,\mathbb{C})$, that leads to the different types of M\"obius transformations mentioned above. In the absence of weak measurements, represented by the non-unitary operator $U_X(i\lambda)$, the matrices associated to the M\"obius transformations that encode the time evolution of coherent states belong to SU$(2)$. As SU$(2)$ is a compact group, the associated M\"obius transformations are always elliptic or parabolic (the trace squared of the associated matrix is bounded by four) and thus only a volume (or subvolume) law phase of entanglement entropy can occur. This hints that entanglement transitions will generically appear in periodic non-unitary circuits where the evolution of the initial state can be encoded in elements of a non-compact Lie group, which may also happen in higher-dimensional systems and in other classes of integrable Floquet systems~\cite{10.21468/SciPostPhys.2.3.021, PhysRevB.103.224303}.

The existence of a MIPT in this class of non-unitary periodic circuits stems from an underlying $\mathcal{PT}$ symmetry of the non-Hermitian Floquet Hamiltonian defined by $U_{\text{F}} = e^{-iH_{\text{F}}}$~\cite{Gopalakrishnan_2021, Kawabata_2023, 10.21468/SciPostPhys.14.5.138, PhysRevResearch.6.013131,PhysRevLett.130.230401, zhou2023entanglement}, 
which has a real spectrum in the volume law phase. The aim of the next sections will be to study the effect of breaking time-translation symmetry, such that no stable Floquet Hamiltonian can be defined, and the dynamics cannot simply be reduced to a quantum quench with a non-Hermitian Hamiltonian. As we shall see, the entanglement transition will be enriched by departing from this effective quantum quench scenario.

\section{Quasiperiodic Fibonacci circuits}
\label{sec:quasiperiodiccircuit}

We now consider a Gaussian circuit (built from the gates~\eqref{unitarygates}) that has no underlying time-translation symmetry, while still following a deterministic non-unitary evolution. To achieve this, we consider a Fibonacci quasiperiodic circuit, in analogy to Fibonacci driven quantum many-body systems~\cite{PhysRevResearch.2.033461, PhysRevResearch.3.023044, PhysRevLett.120.070602}. This circuit is defined by the recursion relation
\begin{equation}
U_n = U_{n-2}U_{n-1}
\end{equation}
given some initial operators $U_0$ and $U_1$. We note that the stroboscopic time defined by the index $n$ corresponds to $F_n$ applied gates, with $F_n$ the $n$-th Fibonacci number.
We choose the initial operators $U_0$, $U_1$ to be given by~\eqref{eq:setpquasi}. We again consider a coherent initial state of the form~\eqref{coherentstate}, such that the time evolution is encoded in products of M\"obius transformations. We denote the SL$(2,\mathbb{C})$ matrix associated with the operators $U_0$ and $U_1$ as $M_0$ and $M_1$, respectively. The evaluation of the time-evolved coherent state thus reduces to the quasiperiodic matrix product
\begin{equation}
    M_n = M_{n-2}M_{n-1},
\label{fibonaccimatrix}
\end{equation}
for each momentum $k$. In particular, the Lyapunov exponent~\eqref{lyapunovexponent} can be defined for the quasiperiodic circuit by approximating it with a periodic circuit whose cycle is composed of the $F_n$ first gates of~\eqref{fibonaccimatrix}. In that case, it may be expressed using~\eqref{lyapunovexponent2} as~\cite{PhysRevResearch.3.023044}
\begin{equation}
\label{eq:usefulforulalyapu}
    \lambda_L = \frac{1}{F_n}\log\left|\frac{|\text{Tr}(M_n)|+\sqrt{|\text{Tr}(M_n)|^2-4}}{2}\right|.
\end{equation}
In particular, the full quasiperiodic circuit will correspond to the limit where the Fibonacci number $F_n$ of individual steps constituting the periodic circuit goes to infinity.
This expression will turn out to be computationally convenient, as the trace of the iterated matrices turns out to be determined from a classical dynamical system.

\subsection{Mapping to quasicrystals}

\begin{figure}[t]
	\includegraphics[width=0.48\textwidth]{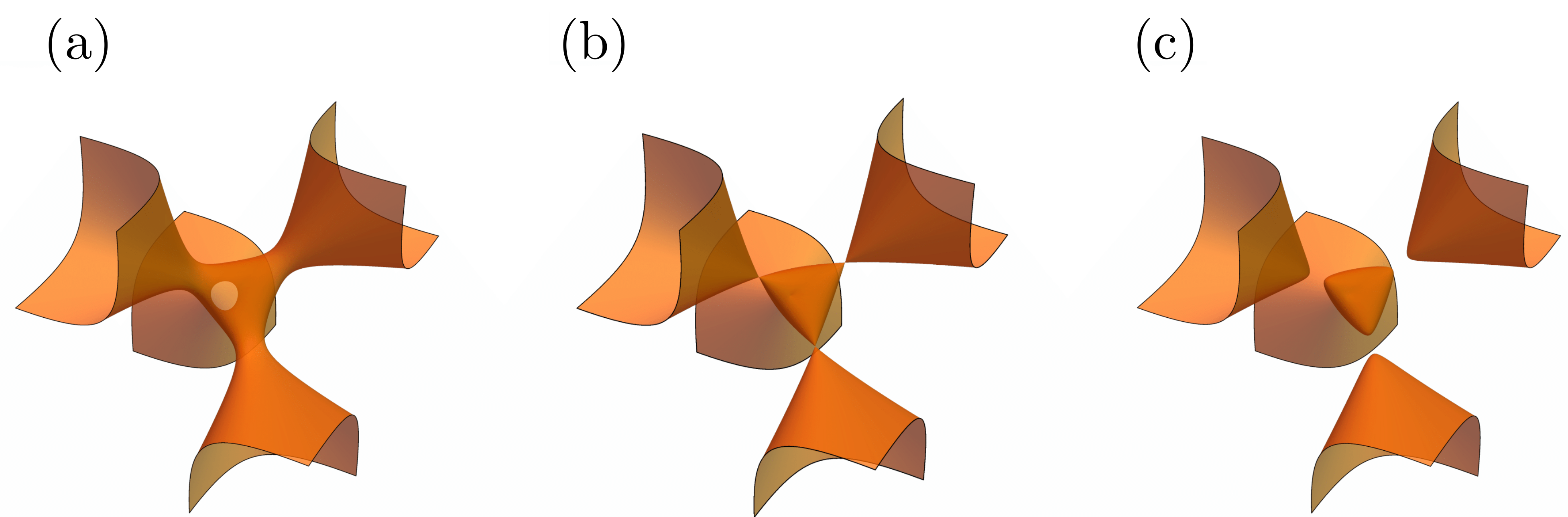}
	\caption{\label{fig:chap4:cayleycubics} Three different topologies for the surface $S_V$ defined by~\eqref{eq:surfaceSV}. (a) For $V>0$ and the manifold is non-compact with the four quadrants connected at the center, (b) For $V=0$, in this case the four quadrants are connected through a single point to the compact manifold at the center, (c) For $-1<V<0$, the manifold $\mathbb{S}_V$ (with the topology of a sphere) in the center is fully disconnected from the four quadrants. The behaviour of the trace map orbits crucially depends on such topology.}
\end{figure}

Although there is no possibility to get a closed form 
of $M_n$ defined by the quasiperiodic product~\eqref{fibonaccimatrix}, we will now make use of a fruitful mapping to Fibonacci quasicrystals in order to make analytical progress. 
Fibonacci quasicrystals are defined by the Schr\"odinger equation
\begin{equation}
\label{schroquasi}
\psi_{n-1}+\psi_{n+1} + \mu V(n)\psi_n=E\psi_n,
\end{equation}
where $V(n)$ is a quasiperiodic potential following 
the
Fibonacci recursion relation. The problem can be translated in terms of transfer matrices by introducing
\begin{equation}
T(n)  = \begin{pmatrix} E-\mu V(n)&-1\\ 1&0 \end{pmatrix}
\end{equation}
such that $T_n = T_{n-2}T_{n-1}$, with $T_n=\prod_{m=1}^{F_n}T(m)$, where the product is taken from right to left. 
In other words, the problem reduces to similar matrix products as~\eqref{fibonaccimatrix}. It was noted that~\eqref{fibonaccimatrix}, together with the condition that the matrices belong to SL$(2,\mathbb{C})$, implies the trace relation~\cite{PhysRevLett.50.1870, Damanik_2016, 1987CMaPh.111..409S, PhysRevLett.57.770}
\begin{equation}
\text{Tr}(M_{n+1})=\text{Tr}(M_n)\text{Tr}(M_{n-1})-\text{Tr}(M_{n-2}).
\end{equation}
Therefore, introducing $x_n=\frac{1}{2}\text{Tr}(M_n)$, this defines a non-linear dynamical system,
\begin{equation}
x_{n+1}=2x_nx_{n-1}-x_{n-2}.
\end{equation}
This provides a numerically efficient way to compute the Lyapunov exponent from~\eqref{eq:usefulforulalyapu}.
For notation convenience, we further introduce $y_n=\frac{1}{2}\text{Tr}(M_{n+1})$ and  $z_n=\frac{1}{2}\text{Tr}(M_{n+2})$. It can be shown for our quasiperiodic circuits that $(x_0, y_0, z_0)\in\mathbb{R}^3$, despite $M_n$ having complex-valued coefficients. Thus, we can define a discrete dynamical map $\mathcal{D}$ on $\mathbb{R}^3$
\begin{equation}
\mathcal{D}:\mathbb{R}^3\rightarrow \mathbb{R}^3,\quad (x_n,y_n,z_n)\mapsto (y_n,z_n,2y_nz_n-x_n).
\label{chap4:fibonaccidynamicslsystem}
\end{equation}
A crucial property of the Fibonacci trace map is that it admits a constant of motion, which reads
\begin{equation}
I(x_n,y_n,z_n)=x_n^2+y_n^2+z_n^2-2x_ny_nz_n-1,
\label{fibonaccitracemapinvarintof}
\end{equation}
and that is the same for any $n\in\mathbb{N}$. It is then natural to consider the level surfaces of the invariant, explicitly given by
\begin{equation}
\label{eq:surfaceSV}
S_V=\{(x,y,z)\in\mathbb{R}^3 \mid I(x,y,z)=V\}.
\end{equation}
The manifold $S_V$ is noncompact for $V>0$ and extends to infinity. For $-1<V<0$, there exists a compact component completely detached from the four quadrants. At $V=0$, the manifold is the Cayley cubic. These different topologies are summarized in Fig.~\ref{fig:chap4:cayleycubics}.
Depending on the sign of the invariant~\eqref{fibonaccitracemapinvarintof}, the typical behaviour of the orbits of the trace map~\eqref{chap4:fibonaccidynamicslsystem} is significantly different. For $V\in(-1,0)$, if the initial point lies on the compact component $\mathbb{S}_V\equiv S_V\cap [-1,1]^3$, the orbit will stay bounded for \textit{infinite} times, as $\mathbb{S}_V$ is invariant under the action of $\mathcal{D}$, i.e., $\mathcal{D}\mathbb{S}_V=\mathbb{S}_V$~\cite{yessen2015newhouse}. As a consequence, the Lyapunov exponent~\eqref{lyapunovexponent} converges to zero, from~\eqref{eq:usefulforulalyapu} in the quasiperiodic limit.
For $V>0$, almost all orbits diverge at infinity with an exponential rate~\cite{Kadanoff1276,PhysRevLett.50.1873}, leading to a nonzero value of the Lyapunov exponent. However, it is known that the set of bounded orbits under the Fibonacci trace map~\eqref{chap4:fibonaccidynamicslsystem} with positive invariant is of measure zero and forms a Cantor set~\cite{yessen2015newhouse, 1992JSP....66..791K, 1987CMaPh.111..409S}. A consequence of this fact is that, even if $V>0$ for all momenta $k$, there may exist a Cantor set of momenta for which the Lyapunov exponent is strictly zero. This is in stark contrast with the periodic case, where the nonconvergent momenta form an interval in the volume law phase, and are $O(1)$ at the phase transition. As we will see later, such a fractal distribution of \textit{isolated} nonconvergent momenta will lead to a critical scaling of the entanglement entropy $S_A(\ell)$.

We now study the orbits of the trace map for our quasiperiodic circuit defined by~\eqref{eq:setpquasi}. The invariant~\eqref{fibonaccitracemapinvarintof} can be computed explicitly in this case, leading to
\begin{align}
\label{transition_quasiperiodicinvariant}
V_k &= \frac{1}{8} \cosh^2(2\lambda) \sin^2(k) \sin^2(4t) \times \nonumber \\
&\quad \left[ 
    -19 + 13 \cosh(4\lambda) 
    + 2 \cosh^2(2\lambda) \right. \nonumber \\
&\quad \left. \times \left(
        3 \cos(4k) 
        - 2 (-4 \cos(4t) + \cos(8t)) \sin^2(2k)
    \right)
\right].
\end{align}
There exists an extended region in the parameter space $(T,\lambda)$ for which $-1<V_k<0$ for an interval of momenta, with initial points $(x_0,y_0,z_0)\in\mathbb{S}_V$. Thus, in this part of the parameter space, there are intervals of momenta with zero Lyapunov exponent, which implies (sub-)volume law scaling of entanglement, as the momenta do not converge to a unique steady state independently of the initial state. Remarkably, this region exactly coincides with the volume law region obtained in the periodic circuit in Sec.~\ref{sec:periodiccircuit} for $U_{\text{F}}=U_0 U_1$. We note that transitions between positive and negative trace map invariants stem from the non-unitarity of our circuit and do not appear in the context of Hermitian Fibonacci quasicrystals, where $I=\frac{\mu^2}{4}$, as can be seen from~\eqref{schroquasi} (in fact, the spectrum of the Hermitian Fibonacci Hamiltonian only consists of critical states, but does not host delocalized states). 
In this sense, we expect a richer crossover between fractal regions with positive invariant and ergodic regions with negative invariant, in which case the trace map has been shown to have classical chaotic properties similar to the standard map~\cite{yessen2015newhouse}.

\begin{figure}[t!]
	\includegraphics[width=0.48\textwidth]{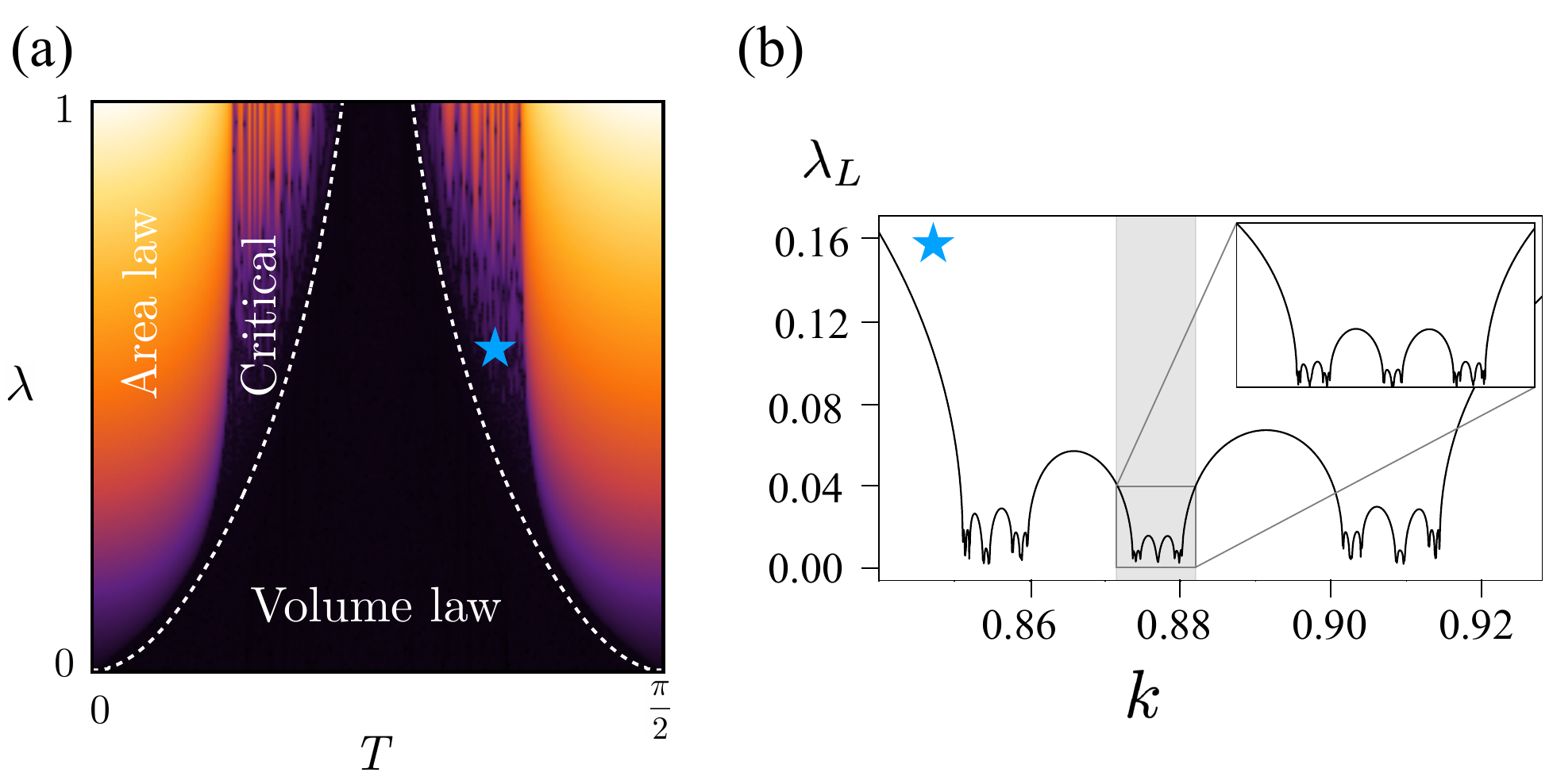}
\caption{\label{fig:phasediamgraquasiperiodic} (a) Phase diagram for the quasiperiodic circuit obtained by minimizing the Lyapunov exponent over momentum space. The emergence of a critical phase is manifest from the fractal structure of the Lyapunov exponent. The white dashed curve corresponds to the phase boundary between critical and volume law phases, as determined analytically from~\eqref{transition_quasiperiodicinvariant}, and coincides with the phase boundary~\eqref{eq:phaseboundwry} of the periodic circuit. We note that although the phase boundary between the area law phase and the critical phase cannot be obtained analytically, it locates asymptotically at $T=\frac{\pi}{8}$ and $T=\frac{3\pi}{8}$ modulo $\frac{\pi}{2}$ in the limit of large $\lambda$. 
(b) The Lyapunov exponent defined as~\eqref{eq:usefulforulalyapu} for parameters $T=\frac{3\pi}{8}-0.1$, $\lambda=0.5$, as a function of the momentum $k$. As can be observed in the inset, there is a clear self-similar structure in the zeros of the Lyapunov exponent, as predicted by the mapping to the Fibonacci quasicrystal.
}
\end{figure}

While the momenta that belong to the Cantor set may not be analytically or numerically accessible due to their fractal nature, there is a simple nontrivial surface in the parameter space $(k,T,\lambda)$ for which the trace map $\mathcal{D}$ is bounded at all Fibonacci times $n$. Taking the measurement rate $\lambda$ to be arbitrary, we consider solutions of the parametric equation
\begin{equation}
\label{parametricequations}
\cos(k)^2 +\cos(4T)\sin(k)=0.
\end{equation}
It is readily shown that for these parameters, $x_0=y_0=0$, i.e., the matrices $M_0$ and $M_1$ are traceless. Thus, the trace map follows the period-six sequence
\begin{equation}
(0,0,z_0)\mapsto (0,z_0,0)\mapsto (z_0,0,0)\mapsto (0,0,-z_0)\mapsto \text{etc}.
\end{equation}
As the trace is bounded for all time steps, it follows that the Lyapunov exponent converges to zero, although the invariant~\eqref{fibonaccitracemapinvarintof} is positive if $z_0>1$. In this specific case, the momenta satisfying~\eqref{parametricequations} converge only for $n$ modulo six. We note that for all $(T,\lambda)\in \left[\frac{\pi}{8},\frac{3\pi}{8}\right]\times \mathbb{R}$, there is at least one momentum $k\in[0,\pi]$ that satisfies~\eqref{parametricequations}. As we will later show, these momenta stay critical even for any random sequence when considering random circuits.

\subsection{Volume law phase}

We first discuss the case where there is an extended interval of momenta for which the trace map invariant is negative. In this case, the Lyapunov exponent vanishes for this interval of momenta, as the trace map $\mathcal{D}$ is constrained to $\mathbb{S}_V$. This region of parameter space $(k,T,\lambda)$ turns out to exactly coincide with the volume law of the periodic circuit discussed in Sec.~\ref{sec:periodiccircuit}. The linear coefficient of the entanglement entropy can be extracted in the periodic case using the fact that stroboscopic time evolution amounts to taking the $n$-th power of an SL$(2,\mathbb{C})$ matrix, and can be dealt with analytically in the infinite time limit. On the other hand, this strategy is not applicable for quasiperiodic circuits, and we find it useful to approximate the quasiperiodic circuit by a periodic circuit, as described previously. We observe that oscillations of $f_n(k)$ for $k\in[k_1,k_2]$ are slower compared to the periodic circuit case, leading to a smaller linear slope of the entanglement entropy. This is expected on physical grounds as the periodic drive has a temporal coherence that maximizes entanglement growth.
We note that outside of the interval of nonconverging momenta $[k_1,k_2]$, there may be momenta that do not converge at infinite times though the associated invariant is positive: 
examples are
the solutions of~\eqref{parametricequations} which also exist in the volume law phase. However, as discussed previously, such momenta form a Cantor set of measure zero, and will only contribute to $O(\log(\ell))$ terms in the entanglement entropy, giving a subleading contribution compared to the momenta from the interval $[k_1,k_2]$. When this interval vanishes and the invariant is positive for all momenta, the scaling of entanglement entropy is dominated by the aforementioned fractal set, leading to a critical phase.

As mentioned previously, the dynamics of the trace map for invariant $V\in(-1,0)$ turns out to be less understood and richer than its counterpart with positive invariant: when the invariant is negative and close to zero, the dynamics of the orbits is generally ergodic and does not display any recurrences: the system is classically chaotic. On the other hand, for values of the invariant close to $-1$, phase space has been shown to be laminated by invariant circles, and the dynamics has recurrences~\cite{yessen2015newhouse}. It will be interesting to further study the link between such classical dynamical features and the volume law scaling of entanglement entropy.

\subsection{Critical phase}

The trace map arguments discussed previously allow us to conclude that there is an extended region in $(T,\lambda)$ space for which a Cantor set of momenta fail to converge in the infinite time limit, as the associated trace map orbit is bounded at all times. This is in stark contrast with the periodic circuit discussed previously, where only a critical line emerges at the transition. We numerically determine the part of the phase diagram for which there is a fractal region by minimizing the Lyapunov exponent over a finite grid of momenta, as plotted in Fig.~\ref{fig:phasediamgraquasiperiodic}(a). 
Although we cannot numerically access the critical momenta that belong to the Cantor set, by continuity of the Lyapunov exponent their neighborhood has
a small Lyapunov exponent in the thermodynamic limit. This enables us to numerically characterize the fractal region in the parameter space. We note that even in the limit of infinitely large measurement strength, $\lambda\rightarrow\infty$, there is still a nonvanishing critical region for $T\in\left[\frac{\pi}{8},\frac{3\pi}{8}\right]$. 
This can be proven using that~\eqref{parametricequations} admits solutions for any $\lambda$, thus there is at least a single critical momentum characterized by a zero Lyapunov exponent.

\begin{figure}[t]
	\includegraphics[width=0.48\textwidth]{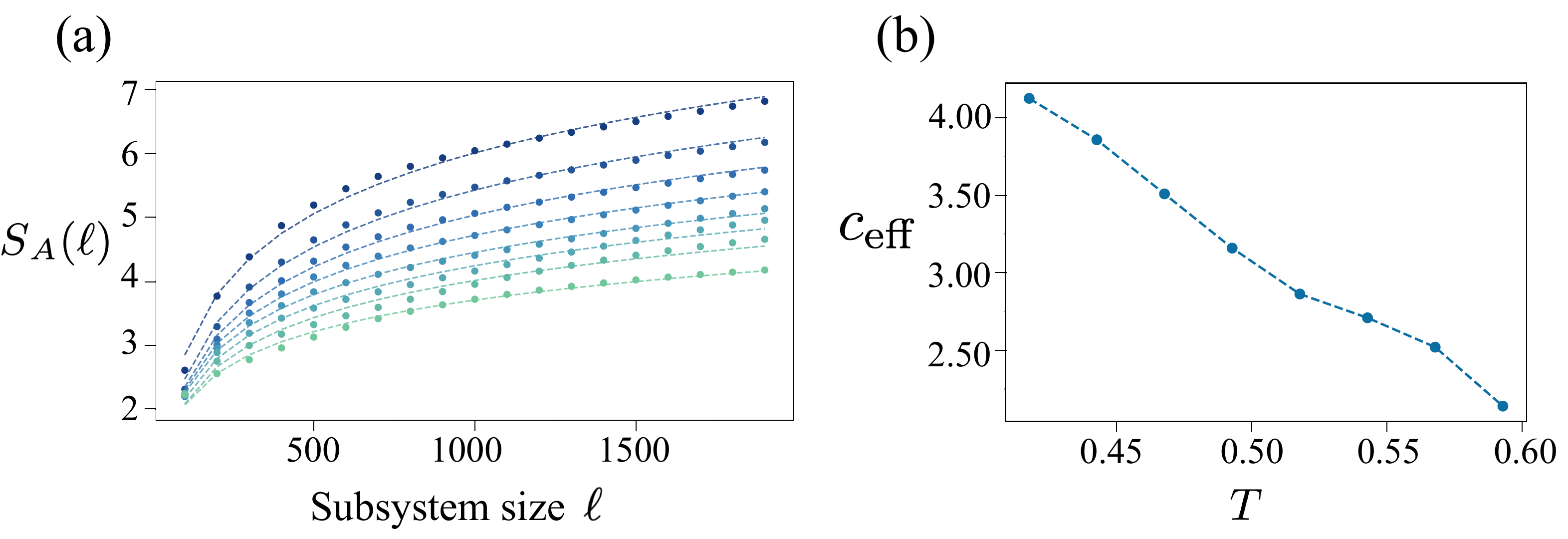}
	\caption{\label{fig:quasientanglemententroysc} (a) Entanglement entropy in the critical phase of the quasiperiodic circuit, with parameters $\lambda=0.8$ and $T=\frac{\pi}{8}+0.025,\frac{\pi}{8}+0.05,\ldots,\frac{\pi}{8}+0.2$ (blue to green), system size $L=10^5$, and $n=24$ Fibonacci steps, which corresponds to order $10^4$ applied gates of the form~\eqref{eq:setpquasi}. The numerically computed entanglement entropy is fitted with $S_{A}\sim \frac{c_{\text{eff}}}{3}\log(\ell)+s_0$ (shown as dashed lines). (b) The effective central charge $c_{\text{eff}}$ varies continuously over the critical phase, here shown for a fixed value of the measurement strength $\lambda=0.8$. 
 }
\end{figure}

As discussed in Sec.~\ref{sec:periodiccircuit}, a measure zero set of nonconvergent momenta leads to a critical scaling of entanglement. However, periodic circuits can only lead to $O(1)$ number of such nonconvergent momenta, while the Fibonacci circuit is characterized by uncountably many of such isolated nonconvergent momenta that are spreading in a fractal way over the momentum space.  Therefore, we expect a nonzero critical scaling of entanglement entropy in this case. We numerically provide evidence for a logarithmic scaling of the entanglement entropy in Fig.~\ref{fig:quasientanglemententroysc}(a), for up to $n=25$ steps, which translates to order $10^5$ individual non-unitary gates. We note that such a numerical computation is subtle in the sense that a finite momentum grid cannot capture the momenta that belong to the fractal set, but we nevertheless observe a convergence of the entanglement entropy by increasing the number of steps $n$, while keeping the order of limits $\ell\ll F_n\ll L$. We provide more details on our numerical approach in App.~\ref{App:numerics}.
The effective central charge is generically nonzero, and varies continuously across the critical phase. As the fractal dimension of the associated Cantor set tends to zero in the $\lambda\rightarrow\infty$~\cite{Damanik_2008}, 
the $O(\log (\ell))$ term vanishes, as expected on physical grounds, given that this corresponds to a limit of infinite measurement strength. On the other hand, we observe that the effective central charge linearly decreases when varying $T$ from the area law phase at $T\approx\frac{\pi}{8}$ to the volume law phase, as clearly seen on Fig.~\ref{fig:quasientanglemententroysc}(b).

\subsection{Area law phase}

The remaining region of the phase diagram, for which the invariant is positive for all momenta and does not contain any nonconvergent momenta, corresponds to the area law region. 
In the case of Fibonacci quasicrystals, there are extended
regions with finite spectral gaps.
Such regions are analogous to the area law phase of  the circuit, where strictly all orbits of the trace map diverge to infinity. 
If an orbit of the trace map is not bounded, it implies that the resulting M\"obius transformation converges to a fixed point, whose value depends on the parity of $n$~\cite{PhysRevResearch.3.023044}.
Thus, in the infinite time limit $n\rightarrow \infty$, all momenta converge to a well-defined steady state, similarly to the area law of the periodic circuit.
The area law region is determined numerically, such that the Lyapunov exponent converges to a finite nonzero value for all momenta. In this case, $f_n(k)$ converges to a different smooth function for $n$ even or odd, as opposed to the periodic circuit. A physical consequence is that the entanglement entropy alternates between two values in the area law. This alternation between two steady states for odd and even Fibonacci times indicates that the circuit behaves as a time quasicrystal, in close analogy to the Fibonacci quasicrystals found in~\cite{Dumitrescu_2018}. 
We can thus define a new stroboscopic time $m=2n$ or $m=2n+1$ for which all momenta converge to a unique steady state, which, applying similar calculations as in Sec.~\ref{sec:periodiccircuit}, leads to an area law of entanglement, $S_A\sim O(1)$.

We stress that even though the area law phase of the quasiperiodic circuit shrunk with respect to the area law phase of the periodic circuit, its interpretation as a purification transition still holds: the final steady state is unique and independent of
the choice of initial coherent state, such that any initial mixed state composed of coherent states purifies in the large time limit (we note that a similar mechanism was recently reported in~\cite{lapierre2025drivennonunitarydynamicsquantum}).

\section{Random circuits}
\label{sec:randomcircuit}

\subsection{General considerations}

After having shown that richer phases could emerge from breaking the discrete time translation symmetry of the periodic circuit down to a quasiperiodic time evolution, it is natural to ask whether random circuits with gates \eqref{unitarygates} may similarly sustain entanglement phase transitions.
We now argue such circuits with initial coherent state~\eqref{coherentstate} 
will generically display transitions from area law to critical scaling of entanglement entropy, but the volume law phase may in general vanish. In order to see this, we stress that the dynamics of the random circuit reduces to random products of SL$(2,\mathbb{C})$ matrices that encode the evolution of fermionic coherent states. In order to characterize this random walk, it is natural to define the Lyapunov exponent as
\begin{equation}
\label{eq:randomlyapu}
\lambda_L=\lim_{n\rightarrow\infty}\frac{1}{n}\mathbb{E}(\log||\Pi_n||),
\end{equation}
where $\Pi_n$ denotes a given length $n$ product of SL$(2,\mathbb{C})$ matrices determined by a random sequence, and $\mathbb{E}(...)$ denotes ensemble average over random realizations. Importantly, Furstenberg's theorem~\cite{10.1214/aoms/1177705909} 
states that random products of SL$(N,\mathbb{R})$ 
matrices almost always have a non-zero Lyapunov exponent (we note that SL$(2,\mathbb{C})$ can be embedded into a higher order real matrix group). This, together with the continuity of the Lyapunov exponent as a function of the parameters $(k,T,\lambda)$~\cite{BOCKER-NETO_VIANA_2017} implies that volume law scaling of entanglement, which necessitates zero Lyapunov over an extended interval of momenta, cannot appear in general. Exceptions to Furstenberg's theorem typically form a measure zero set in parameter space~\cite{10.21468/SciPostPhys.13.4.082}, and would therefore at most lead to a critical phase.

We note that the statements about the absence of a volume law phase for the circuits~\eqref{unitarygates} may be reinterpreted in terms of localization transitions in disordered one-dimensional lattice models, by noting that the transfer matrix in this context plays a similar role as the M\"obius transformation in the present work.
Indeed, it is well-known that for Hermitian disordered lattice models in one dimension, Anderson localization precludes the existence of delocalized states. This can be understood from Furstenberg's theorem because the Lyapunov exponent associated with the random products of transfer matrices is (almost always) nonzero.
We will now explore in the rest of the section random non-unitary circuits that can sustain volume law phases thanks to an underlying compactness of the group manifold that encodes time evolution.

\subsection{Random circuits with volume law phases}
\label{sec:randomcircuitsvolume}

A crucial assumption needed for Furstenberg's theorem to apply is that the smallest subgroup containing the random walk generated by $M_0$ and $M_1$ is noncompact. An important counterexample is when the matrices $M_0,M_1\in \text{SU}(2)$, in which case the random product they generate is of course an element of SU$(2)$, and thus the Lyapunov exponent is zero. This is what happens for the unitary random circuit in the absence of weak measurements, $\lambda=0$, in which case only a volume law emerges. Nevertheless, whenever $\lambda\neq0$ the matrices are general elements of SL$(2,\mathbb{C})$ and therefore \textit{a priori} generate a sequence that is not contained in a compact subgroup.
We now discuss two protocols that lead to a robust volume law at finite measurement strength $\lambda$ and escape Furstenberg's theorem because the dynamics is effectively constrained to SU$(2)$.

\subsubsection{Protocol I: Random time evolution} 
Let us first consider the example considered in Secs.~\ref{sec:periodiccircuit} and~\ref{sec:quasiperiodiccircuit}, where the two building blocks of the circuit are the operators $U_0$ and $U_1$ in~\eqref{eq:setpquasi}. As noted previously, the volume law phases realized in these circuits are of the same kind
and one would thus expect them to survive even for random circuits. We will now argue that indeed the volume law phase determined analytically by~\eqref{eq:phaseboundwry} survives for $\textit{any}$ random sequences of $U_0$ and $U_1$. 
In order to make progress, let us note that the associated matrices $M_0$ and $M_1$ that encode time evolution of coherent states do not lie in SU$(2)$ as long as $\lambda\neq0$. Nevertheless, the two matrices are related by the conjugation symmetry
\begin{equation}
\label{eq:relationmatrices}
M_1 = \sigma_z M_0^*\sigma_z.
\end{equation}
Furthermore, as noted in the previous sections, both matrices have real trace, and in the $(T,\lambda)$ region delimited by~\eqref{eq:phaseboundwry} there is always a nonvanishing interval of momenta $k$ for which $|\text{Tr}(M_0M_1)|<2$. As we show in App.~\ref{app:proofsimilarity}, these conditions imply that the two matrices $M_0$ and $M_1$ can be simultaneously brought into SU$(2)$ matrices via a similarity transformation $S$. Therefore, any product $\Pi_n$ of length $n$ of the two matrices reads
\begin{equation}
\Pi_n = S \Sigma_n S^{-1}, 
\end{equation}
where $\Sigma_n\in\text{SU}(2)$. It is now clear that the Lyapunov exponent, as defined in~\eqref{eq:randomlyapu}, necessarily vanishes because the infinitely large string of operators effectively reduces to an SU$(2)$ transformation (up to the boundary terms), which has a bounded norm.
Consequently, the entanglement entropy $S_A(\ell)$ grows with subsystem size $\ell$ when the above conditions are fulfilled, following the arguments discussed previously. 
This provides us with a bigger picture for the robust volume law phase observed in the periodic and quasiperiodic circuits: the volume law phase exists because the dynamics generated by the gates $U_0$ and $U_1$ on coherent states is restricted to a compact manifold.

It is now natural to ask whether there could be a critical phase outside the volume law region, similarly to the Fibonacci circuit in Sec.~\ref{sec:quasiperiodiccircuit}. As explained previously, when the two matrices cannot be simultaneously brought into SU$(2)$ matrices, Furstenberg's theorem will in general apply. Nevertheless, there may still be a measure zero set of parameters $(k,T,\lambda)$ with zero Lyapunov exponent.
An example is the case $\text{Tr}(M_0)=\text{Tr}(M_1)=0$. This tracelessness condition implies that $M_0^2=M_1^2=-\mathbb{I}$, which in turn leads to a zero Lyapunov exponent~\cite{10.21468/SciPostPhys.13.4.082}. This follows from the square root scaling
\begin{equation}
\label{eq:zeolyapunovexponent}
\mathbb{E}(\log||\Pi_n||)\sim\sqrt{n},
\end{equation}
 which can be derived using that the matrix norm of $\Pi_n$ for generic sequences is `reduced' by $M_0^2=M_1^2=-\mathbb{I}$.
 In the case of~\eqref{eq:setpquasi}, the tracelessness condition is equivalent to Eq.~\eqref{parametricequations}.
We conclude that for $T\in\left[(2m+1)\frac{\pi}{8},(2m+1)\frac{\pi}{4}\right]$ and $\lambda$ arbitrary, there always exists at least one momentum $k$ for which the Lyapunov exponent is strictly zero in the infinite time limit. Similarly to the phase transition of the periodic circuit case (see Sec.~\ref{sec:periodiccircuit}), the existence of a single momentum with zero Lyapunov exponent generically implies $O(\log(\ell))$ scaling of $S_{A}(\ell)$, with zero effective central charge.

\subsubsection{Protocol II: Random measurement sign}

The way weak measurements are implemented in the circuit defined in Sec.~\ref{sec:setup} is through the non-unitary operator $U_{X}(i\lambda)$, which models measuring ancilla qubits coupled to the spin chain at each site with strength $\lambda$, and postselecting the outcome. For potential experimental realizations of such Gaussian circuits, it is desirable to understand the effect of removing the postselection. 
This would require taking the sign of $\lambda$ randomly at each site and each time step, and would directly break both spatial and temporal translation symmetry. 
Instead, we insist on maintaining spatial translation invariance, as it is necessary for our analytical treatment of the circuit, and we consider the sign of $\lambda$ to change randomly at each time step. An example of such a circuit is generated by alternating randomly between gates
\begin{align}
\label{eq:setpquasinew}
U_+ &= U_{ZZ}(-T)U_X(i\lambda)U_{YY}(T),\\
U_- &= U_{ZZ}(-T)U_X(-i\lambda)U_{YY}(T).\nonumber
\end{align}
The gates $U_+$ and $U_-$ are designed in such a way that the associated matrices $M_+$ and $M_-$ follow the conjugation symmetry
\begin{equation}
\label{eq:relationmatrices2}
M_+ = \sigma_x M_-^*\sigma_x.
\end{equation}
This, together with the reality condition of their traces, as well as $|\text{Tr}(M_-M_+)|<2$, implies that a volume law emerges. In this case, for all $(T,\lambda)$ there exists an interval of momenta for which $M_-$ and $M_+$ are simultaneously transformable to SU$(2)$ transformations. Therefore, there is only a volume law phase, and no phase transition to the area law subsists.
It is an interesting future direction to design random non-unitary circuits from the gates~\eqref{unitarygates}, for which the sign of $\lambda$ is randomly chosen and such that a phase transition still emerges.

\subsection{Random multipolar circuits}
\label{sec:randomdipolarcircuits}

In Sec.~\ref{sec:randomcircuitsvolume} we have demonstrated that a volume-to-area law phase transition can exist in special non-unitary random circuits built from \eqref{eq:setpquasi}. In general, the set of gates which satisfy the criteria for the existence of the volume phase may be limited. We now design a larger set of random circuits which can also show transitions between area and volume law phases under some minimal extra assumption on the randomness of the circuit. Specifically, we consider random dipolar circuits, which are inspired by the random dipolar driven quantum many-body systems~\cite{PhysRevLett.126.040601, liu2025prethermalizationrandommultipolardriving, mo2025complex}.

For concreteness, we now consider the gates introduced in~\cite{PhysRevLett.130.230401}, which were used to demonstrate a volume-to-area law transition for periodic circuits. The two layers are
\begin{align}
\label{eq:setpquasinewew}
U_+ &= U_X(i\lambda)U_X\left(\frac{\pi}{4}-T\right)U_{ZZ}\left(\frac{\pi}{4}-T\right),\\
U_- &= U_X(i\lambda)U_X\left(\frac{\pi}{4}+T\right)U_{ZZ}\left(\frac{\pi}{4}+T\right),\nonumber
\end{align}
and thus the associated matrices $M_+$ and $M_-$ satisfy the conjugation symmetry $M_+ = \sigma_z M_-^*\sigma_z$.
However, $M_+$ and $M_-$ have complex-valued traces, $\text{Im}(\text{Tr}(M_+)), \text{Im}(\text{Tr}(M_-))\neq0$. This violates one of the assumptions from Sec.~\ref{sec:randomcircuitsvolume}, and therefore the volume law phase cannot exist for a general random circuit alternating between $U_+$ and $U_-$ with probability $\frac{1}{2}$.
Nevertheless, we may instead consider a dipolar random circuit by defining the `dipoles'
\begin{align}
\label{eq:setrandomdipolar}
\tilde{U}_+ &= U_+U_-,\\
\tilde{U}_- &= U_-U_+,\nonumber
\end{align}
and consider a random sequence alternating between $\tilde{U}_+$ and $\tilde{U}_-$ with probability $\frac{1}{2}$.
It is straightforward to show that $\text{Tr}(M_+M_-)\in\mathbb{R}$, such that the new building blocks $\tilde{M}_+=M_+M_-$ and $\tilde{M}_-=M_-M_+$ satisfy all the criteria to be simultaneously transformed into SU(2) matrices for a nonzero interval of momenta as long as the condition
\begin{equation}
\lambda< \frac{1}{2}\text{arcsinh}\left[\frac{\cos^2(2T)}{\sin(2T)}\right]
\label{eq:arcsinhphasediagram}
\end{equation}
is fulfilled. Consequently, a (sub)volume law phase exists for any random sequence alternating between $\tilde{U}_+$ and $\tilde{U}_-$, which in turn imposes a dipolar random sequence on the operators $U_+$ and $U_-$.

\begin{figure}[t]
	\includegraphics[width=0.48\textwidth]{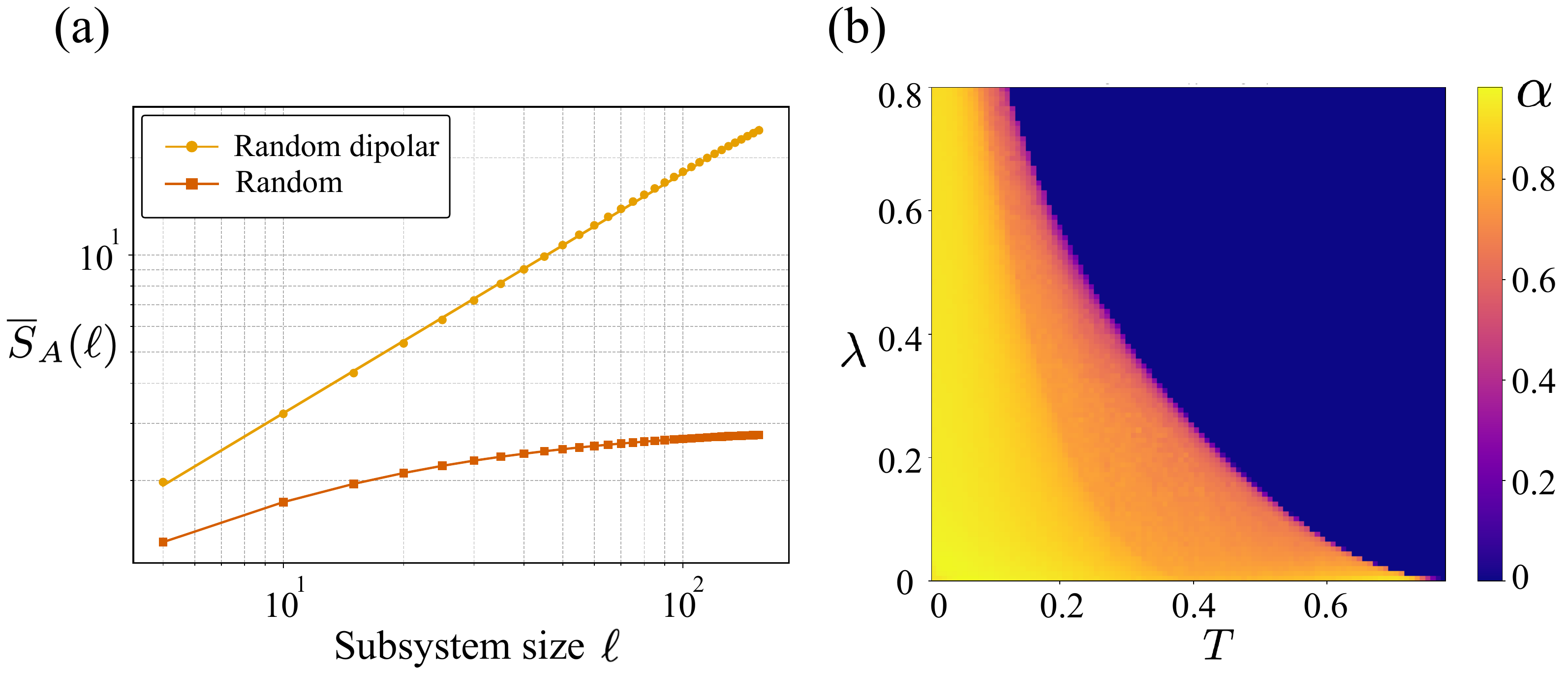}
	\caption{\label{fig:randomcomparison} (a) Scaling of the averaged entanglement entropy $\overline{S}_A(\ell)$ for the circuit alternating randomly between gates $U_+$ and $U_-$ in~\eqref{eq:setpquasinewew} (orange) and for the dipolar random circuit alternating between the gates~\eqref{eq:setrandomdipolar} randomly (yellow), averaged over 1000 random realizations, with parameters $T=\frac{\pi}{10}$, $\lambda=0.2$, total system size $6000$ and for $n=4000$ random steps. As is manifest from the plot, entanglement entropy scales as a power law in the random dipolar case in the subvolume law phase, $\overline{S}_A(\ell)\sim \ell^{\alpha}$ with $\alpha\approx 0.74$ (as opposed to linear scaling for the periodic circuit). For purely random circuits, there is no extensive growth of entanglement entropy in the thermodynamic limit, and $\overline{S}_A(\ell)$ converges to a constant independent of subsystem size for $\ell$ large enough. (b) Scaling exponent $\alpha$ for the random dipolar circuit in the subvolume law phase, for 20 disorder realizations, system size $4000$ and $n=3000$.
 }
\end{figure}

We numerically compare in Fig.\ \ref{fig:randomcomparison}(a) the scaling of the averaged entanglement entropy for both the random circuit and the random dipolar circuit~\eqref{eq:setrandomdipolar} built from the gates~\eqref{eq:setpquasinewew}. We find that for the random circuit, the entanglement entropy converges to a finite value such that $\overline{S}_A(\ell)\sim O(1)$ in the thermodynamic limit, even when the condition~\eqref{eq:arcsinhphasediagram} is fulfilled. This can be explained by Furstenberg's theorem, as the Lyapunov exponent $\lambda_L$ is always strictly positive.
In contrast, by adding some additional structure on the randomness and considering a random dipolar evolution, we numerically find that the entanglement entropy scales sublinearly, $\overline{S}_A(\ell)\sim \ell^{\alpha}$. This has to be contrasted with the volume law phase found in Floquet circuits, for which the entanglement entropy scaling can be shown to be strictly linear using Szeg\"o’s lemma~\cite{PhysRevLett.130.230401}. In particular, we observe that the coefficient of the power law scaling of entanglement entropy varies continuously in the subvolume law phase, as seen in Fig.\ \ref{fig:randomcomparison}(b).

The above construction can readily be applied to random multipolar circuits of order $n$, for which the volume law phase can be shown to be the same as for the dipolar case $n=2$. In particular, it is known that in the limit $n\rightarrow\infty$, the evolution follows the Thue-Morse quasiperiodic sequence~\cite{PhysRevLett.126.040601}, whose first steps read
\begin{equation}
\label{Eq:Thue-Morse}
U_+, \quad U_+U_-,\quad  U_+U_-U_-U_+,
\end{equation}
and the next steps are built by replacing $U_+$ by $U_+U_-$ and $U_-$ by $U_-U_+$.
Similarly to the Fibonacci circuit studied in Sec. 
\ref{sec:quasiperiodiccircuit}, the trace of the quasiperiodic product of SL$(2,\mathbb{C})$ matrices can be mapped onto a classical dynamical system, as detailed in App.~\ref{app:tracemapss}. In particular, the trace map leads to three distinct possible behaviors depending on the initial parameters:  
(i) an extended region with zero Lyapunov exponent, (ii) an extended region with strictly positive Lyapunov exponent, and (iii) an extended region with strictly positive Lyapunov exponent coexisting with a measure zero set of bounded orbits. The case (i) maps to the volume law discussed previously, and the case (ii) maps to an area law phase. As discussed for the Fibonacci circuit, the case (iii) could lead to a critical (log-law) region in parameter space. Nevertheless, as shown in App.~\ref{app:tracemapss}, the region (iii) is inaccessible for our parameter space.
This implies that for the quasiperiodic Thue-Morse circuit, there are only extended volume and area law phases, but no critical phase survives, as opposed to the Fibonacci circuit.

\section{Outlook}

In this work, we have investigated the entanglement dynamics of a class of non-unitary Gaussian circuits that lack translation invariance in time, both through quasiperiodic as well as random evolutions.

While on general grounds the entanglement transitions discussed in this work may be expected to vanish without discrete time-translation invariance, we have demonstrated that richer entanglement phases may emerge beyond the Floquet circuits.
In fact, while the area and volume law phases in the periodic circuit are separated by a critical phase boundary, an extended critical phase emerges from quasiperiodicity. The existence of this critical phase stems from the fractality of Fibonacci evolution, leading to a Cantor set of initial momenta that never converge under time evolution.
On the other hand, for fully random circuits, we have shown that the naive expectation that Furstenberg's theorem should preclude the existence of a volume law phase may not always hold, and we explicitly designed random non-unitary circuits that sustain area-to-(sub)volume law transitions thanks to an emergent SU(2) dynamics.

The entanglement phases emerging in both the quasiperiodic and random non-unitary circuits can be linked to localization transitions in quasiperiodic and disordered lattice models. In fact, the log-law phase of the Fibonacci circuit maps to the multifractal states of the 1D Fibonacci quasicrystal~\cite{RevModPhys.93.045001, Damanik_2016}, while the volume-to-area law transition for the dipolar random circuit is analogous to a recently found localization transition in non-Hermitian disordered lattice models~\cite{PhysRevB.111.235412}. It would be interesting to understand further this mapping to localization transitions in 1D models. Furthermore, recent studies found a link between MIPT for free fermions in $(d+1)$ dimensions and localization transitions in $d+1$ spatial dimensions~\cite{PRXQuantum.2.040319, PhysRevX.12.011045, PhysRevX.13.041045, PhysRevX.13.041046,PhysRevLett.132.110403, chahine2023entanglement}. It would be particularly fruitful to understand how our results fit into this general picture, and in particular to understand whether the transitions found in this paper can be reinterpreted via spacetime duality~\cite{PRXQuantum.2.040319}.

In all of our discussions, the dynamics of the system was confined to the subspace of fermionic coherent states during non-unitary time evolution, and could thus simply be encoded by SL$(2,\mathbb{C})$ transformations. It would be interesting to understand whether leaving such a subspace would still lead to entanglement transitions, despite the dynamics not being encoded in a classical dynamical system anymore.
Furthermore, it is desirable to design similar classes of integrable non-unitary circuits in higher dimensions. Higher-dimensional systems may involve a larger algebra than SL$(2,\mathbb{C})$, which could lead to richer entanglement transitions; we expect such transitions to happen whenever the associated group is non-compact.

\section{Acknowledgments}

B.L.~thanks Dan~S.~Borgnia, Vir~B.~Bulchandani, Etienne~Granet, Anish Kulkarni, Marco Schiro,  Martina~O.~Soldini, William~Witczak-Krempa and Carolyn~Zhang for stimulating discussions. B.L.~acknowledges financial support from the Swiss National Science Foundation (Postdoc.Mobility Grant No. 214461). 
S.R. is supported by a Simons
Investigator Grant from the Simons Foundation (Award No. 566116).

\appendix

\section{Numerical evaluation of $S_A(\ell)$ for the Fibonacci circuit}
\label{App:numerics}

In this appendix, we provide more detail on the numerical simulations shown in Fig.~\ref{fig:quasientanglemententroysc} for the Fibonacci non-unitary circuit in the critical (fractal) phase. 
An important limitation in computing the entanglement entropy numerically is that the set of critical momenta in $\tilde{f}_n(k)$ at large Fibonacci times $n$ forms a Cantor set of zero measure. Nevertheless, the logarithmic scaling of entanglement entropy in the steady state ultimately comes from this set of critical momenta. This leads to a challenging issue that in numerical calculations any finite momentum grid will have probability zero to contain the (infinitely many) critical momenta that belong to the Cantor set, apart from the period-6 critical momenta satisfying~\eqref{parametricequations}.

In order to optimize the numerical evaluation of the entanglement entropy for given parameters $(T,\lambda)$, we compute the $n$-th iteration of the trace map~\eqref{chap4:fibonaccidynamicslsystem} for all the discrete momenta $k$, up to Fibonacci step $n=24$, which corresponds to order $10^4$ non-unitary gates. While for all momenta in the finite grid there is a probability 1 that the orbit of the trace map will diverge to infinity, the momenta closer to the Cantor set will have a higher escape time, because their Lyapunov exponent is closer to zero (but importantly nonzero). We therefore numerically approximate the escape time for each momentum $k\in\Lambda_+$ as the first Fibonacci step $N_k$ such that the map~\eqref{chap4:fibonaccidynamicslsystem} exceeds $10^7$. This allows us to optimize the numerical evaluation of entanglement entropy by considering a high number of iterations for the momenta near the fractal set. The Fourier coefficients~\eqref{fouriercoeffcieitnseq} are then numerically evaluated using fast Fourier transform, and will typically have a long-range decay in the fractal phase, leading to the logarithmic scaling of entanglement entropy shown in Fig.~\ref{fig:quasientanglemententroysc}(a).

\section{Proof of the volume law phase and phase transition for random circuits}
\label{app:proofsimilarity}

In this appendix, we establish the conditions under which a volume law entanglement phase emerges from the non-unitary evolution generated by two alternating gates $U_+, U_-$ built out of the operators~\eqref{unitarygates}, which act on coherent states~\eqref{coherentstate} with $M_+, M_- \in \mathrm{SL}(2, \mathbb{C})$.  These matrices may represent driving protocols such as those in Eqs.~\eqref{eq:setpquasi},~\eqref{eq:setpquasinew} and~\eqref{eq:setpquasinewew}.

Specifically, we consider the case where the following conditions are satisfied:
\begin{enumerate}
    \item The matrices obey a conjugation symmetry: $M_+ = \sigma_i M_-^* \sigma_i$, where $\sigma_i$ are Pauli matrices ($i = 0, x, y, z$).
    \item The traces of both matrices are real: $\mathrm{Im}[\mathrm{Tr}(M_{\pm})] = 0$.
    \item The trace norms are bounded by $2$: $|\mathrm{Tr}(M_{\pm})| \le 2$ and the combined trace satisfies $\mathrm{Tr}(M_+ M_-) \le 2$.
\end{enumerate}

Under these assumptions, both $M_+$ and $M_-$ can be simultaneously transformed into $\mathrm{SU}(2)$ matrices via similarity transformations. Consequently, the resulting non-unitary dynamics support a volume-law entangled phase in the infinite time limit. This is the case because $\mathrm{SU}(2)$ is a compact Lie group, and therefore any arbitrary sequence formed by these matrices must have a zero Lyapunov exponent. This argument therefore holds for the periodic, quasiperiodic, and random circuits studied in this work.

Each of the three gate constructions we consider automatically satisfies the first condition.  
In particular, for both $U_0$ and $U_1$ in Eq.~\eqref{eq:setpquasi}, as well as in~\eqref{eq:setpquasinewew}, the gates obey
\begin{align}
\label{Case:sigz}
    M_+ = \sigma_z M_-^* \sigma_z.
\end{align}
On the other hand, the gates in Eq.~\eqref{eq:setpquasinew} satisfy
\begin{align}
\label{Case:sigx}
    M_+ = \sigma_x M_-^* \sigma_x.
\end{align}

In the following, we focus on the case defined by Eq.~\eqref{Case:sigx}, while noting that the derivation naturally generalizes to other choices of $\sigma_i$.

\begin{proof}
    Conditions 1 and 2 imply that the eigenvalues of both $M_+$ and $M_-$ take the form $u \pm iv$, where $u^2 + v^2 = 1$. Therefore, these matrices can be diagonalized as
\begin{align}
M_+ &= P \Lambda P^{-1}, \quad 
P = \begin{pmatrix} a & b \\ c & d \end{pmatrix}, \quad
\Lambda = \begin{pmatrix} u+iv & 0 \\ 0 & u-iv \end{pmatrix}, \notag \\
M_- &= \sigma_x P^* \Lambda^* (P^*)^{-1} \sigma_x,
\end{align}
where each column of $P$ is an eigenvector of $M_+$.
For simplicity, we assume that $P$ is normalized and satisfies $\det(P) = 1$.
By construction, the matrix $P$ diagonalizes $M_+$ into $\Lambda$, which belongs to the group $\mathrm{SU}(2)$. However, when applying the same transformation to $M_-$, we obtain
\begin{align}
    P^{-1} M_- P = P^{-1} \sigma_x P^* \Lambda^* (P^*)^{-1} \sigma_x P,
\end{align}
which, in general, does not lie within $\mathrm{SU}(2)$. To understand its structure, we examine the matrix 
\begin{align}
    \frac{P^{-1} \sigma_x P^*}{\sqrt{\det(\sigma_x)}} =
    \begin{pmatrix}
    i(a^* b - c^* d) & i(|b|^2 - |d|^2) \\
    -i(|a|^2 - |c|^2) & -i(a b^* - c d^*)
\end{pmatrix},
    \label{Eq:normalsimi}
\end{align}
where we normalized the equation.
It is worth noting that the diagonal elements of this matrix are complex conjugates of each other, as needed for $\mathrm{SU}(2)$ matrices. However, the off-diagonal elements generally lack this structure and can deviate from the constraints imposed by $\mathrm{SU}(2)$, indicating that the transformed matrix in Eq.~\eqref{Eq:normalsimi} does not, in general, remain within the $\mathrm{SU}(2)$ group.
We can nonetheless perform a similarity transformation with the matrix
\begin{align}
    W = \begin{pmatrix}
1 & 0 \\
0 & \sqrt{-\dfrac{|b|^2 - |d|^2}{|a|^2 - |c|^2}}
\end{pmatrix},
\end{align}
on Eq.~\eqref{Eq:normalsimi}, such that we obtain
\begin{align}
U &= \frac{1}{\sqrt{\det(\sigma_x)}}\, W  P^{-1} \sigma_x P^*  W^{-1} 
\nonumber \\
  &=  \begin{pmatrix}
    i(a^* b - c^* d) & -i\sqrt{\Delta} \\
-i\sqrt{\Delta} & -i(a b^* - c d^*)
\end{pmatrix},
\end{align}
with $\Delta=(|a|^2 - |c|^2)( |d|^2-|b|^2)$.
Thus, $U$ becomes an $\mathrm{SU}(2)$ matrix whenever $\mathrm{Re}[\Delta] \ge 0$ and $\mathrm{Im}[\Delta]=0$.
Further analysis shows that this is equivalent to requiring
\begin{align}
    &\mathrm{Re}[\mathrm{Tr}(M_+ M_-)] \leq 2,\\
& \mathrm{Im}[\mathrm{Tr}(M_+ M_-)] =0,
\end{align}
because of the identity
\begin{align}
\mathrm{Tr}(M_+ M_-)
&= 2 + 4(|a|^2 - |c|^2)(|b|^2 - |d|^2)
.
\end{align}
It is straightforward to verify that the first condition ensures that $\mathrm{Tr}(M_+ M_-)$ is real. Moreover, numerical evidence shows that the inequality $|\mathrm{Tr}(M_{\pm})| \le 2$ is automatically satisfied whenever
\begin{align}
\label{Eq:condsu2}
    \mathrm{Tr}(M_+ M_-) \leq 2
\end{align}
holds. 

In conclusion, the combination of condition~1 and Eq.~\eqref{Eq:condsu2} is sufficient to ensure that the effective dynamics is governed by an emergent $\mathrm{SU}(2)$ structure, leading to a zero Lyapunov exponent for any sequence.

\end{proof}

\section{Trace map and invariant areas for Thue-Morse circuits}
\label{app:tracemapss}

In the following, we show that our analysis for the random dipolar circuits can be naturally generalized to the Thue–Morse sequence defined in Eq.~\eqref{Eq:Thue-Morse}. In this case, the trace map that encodes the trace of the iterated matrices $M_n$ takes the form
\begin{align}
    x_{n+1} = x_{n-1}^2 (x_n - 2) + 2, \quad n \ge 2,
    \label{Eq:TM}
\end{align}
with $x_n \equiv \mathrm{Tr}(M_n)$.
This recurrence defines a nonlinear dynamical system whose behavior can be understood by identifying invariant regions in trace space. 

To facilitate the analysis, we introduce new variables $p_n = x_n^2$ and $q_n = x_{n+1}$, converting Eq.~\eqref{Eq:TM} into a two-dimensional discrete dynamical map:
\begin{equation}
\mathcal{K}: (p, q)\mapsto (q^2, pq - 2p + 2).
\label{Eq:tracemap2d}
\end{equation}
The $(p, q)$ plane can then be partitioned into three invariant regions, each of them being stable under the application of $\mathcal{K}$:
\begin{itemize}
    \item[-] {Region I}: $\{(p, q) \mid p \ge 0,\; p - 2 \le q \le 2\}$,
    \item[-] {Region II}: $\{(p, q) \mid p \ge 0,\; q \ge 2\}$,
    \item[-] {Region III}: $\{(p, q) \mid p \ge 0,\; q \le 2,\; q \le p - 2\}$.
\end{itemize}
Region I leads to compact and bounded classical dynamics, implying zero Lyapunov exponent, and thus corresponds to a (sub)volume law phase. In contrast, both Region II and Region III lead to non-compact dynamics; Region II and almost all points in Region III escape to infinity under iteration of the map~\eqref{Eq:tracemap2d}, leading to strictly positive Lyapunov exponents and area-law entanglement entropy. However, Region III distinguishes itself from Region II by its intricate fractal boundaries that contain sets of non-escaping points, which can be mapped to a fixed point similar to the Fibonacci trace map~\eqref{chap4:fibonaccidynamicslsystem} discussed in Sec.~\ref{sec:quasiperiodiccircuit}.

For the non-unitary circuits discussed in the present work, only Regions I and II are dynamically accessible, and Region III is inaccessible. This can be seen from the condition defining Region III, which requires $q_1 \le 2$. This in turn implies Eq.~\eqref{Eq:condsu2}, which characterizes gates that can be simultaneously transformed into $\mathrm{SU}(2)$ and hence fall within Region I.  Notice that $q_1\le 2$ is also a requisite for Region III, but the conditions directly lead to Region I due to the compactness of $\mathrm{SU}(2)$.
Therefore, Region III does not appear in our phase diagram. The direct physical consequence is that there is no fractal log-law phase for the Thue-Morse circuit.

\bibliography{ref}
\end{document}